\documentclass[12pt]{JHEP3}
\usepackage{amsmath,amssymb,epsfig,graphicx}
\usepackage{url}

\newcommand{\be}{\begin{equation}}
\newcommand{\ee}{\end{equation}}
\newcommand{\beqa}{\begin{eqnarray}}
\newcommand{\eeqa}{\end{eqnarray}}

\def\eeq{\end{equation}}

\relax

\def\beq{\begin{equation}}
\def\eeq{\end{equation}}
\def\bea{\begin{eqnarray}}
\def\eea{\end{eqnarray}}

\title{Q-PYTHIA: a medium-modified  implementation of final state radiation}

\author{N\'estor~Armesto${}^{\,a}$,
Leticia Cunqueiro${}^{\,b}$
and Carlos~A.~Salgado${}^{\,a}$\\
\vspace{0.1in}

${}^{\,a}$Departamento de F\'\i sica de Part\'\i culas and IGFAE,
Universidade de Santiago de Compostela, E-15706 Santiago de Compostela, Galicia--Spain
\vspace{0.1in}

${}^{\,b}$ Istituto Nazionale di Fisica Nucleare, Laboratori Nazionali di Frascati,  I-00044 Frascati (Roma), Italy
\vspace{0.1in}

E-mail addresses: {\tt nestor.armesto@usc.es,
Leticia.Cunqueiro.Mendez@cern.ch, carlos.salgado@usc.es}
}

\abstract{
We present a Monte Carlo implementation, within PYTHIA, of medium-induced gluon radiation in the final state branching process. Medium effects are introduced through an additive term in the splitting functions computed in the multiple-soft scattering approximation. 
The observable effects of this modification are studied for different quantities as fragmentation functions and the hump-backed plateau, and transverse momentum and angular distributions. The anticipated increase of intra-jet multiplicities, energy loss of the leading particle and jet broadening are observed as well as modifications of naive expectations based solely on analytical calculations. This shows the adequacy of a Monte Carlo simulator for jet analyses. Effects of hadronization are found to wash out medium effects in the soft region, while the main features remain.  To show the performance of the implementation and the feasibility of our approach in realistic experimental situations  we provide some examples: fragmentation functions, nuclear suppression factors, jet shapes and jet multiplicities. The package containing the modified routines is available for public use. This code, which is not an official PYTHIA release, is called Q-PYTHIA. We also include a short manual to perform the simulations of jet quenching.}

\keywords{Jet Quenching; Medium-Induced Gluon Radiation}
\preprint{\today}

\begin{document}

\section{Introduction}
\label{intro}

Two of the most striking experimental observations at the Relativistic Heavy Ion Collider (RHIC) at Brookhaven National Laboratory \cite{rhic} are: The suppression of particles with large transverse momentum produced in nucleus-nucleus collisions compared with the expectations from an incoherent superposition of nucleon-nucleon ones, and; The disappearance or suppression of the yield of high transverse momentum particles in the region in azimuth opposite to a high transverse momentum one. These two phenomena are manifestations of medium effects which are usually known under the generic name of  {\it jet quenching}.

The most common explanation for jet quenching is given in terms of radiative energy loss of a high-energy parton traversing a dense partonic medium, see the reviews \cite{rel}. Within this framework, the amount of energy loss measured allows to map medium properties as energy density, temperature, etc. Models implementing these ideas have been successful in describing the mentioned phenomena for light hadrons, while problems persist for heavy flavors, see e.g. \cite{Armesto:2005mz}.

In spite of  these phenomenological successes, both the measurements and their theoretical description in terms of radiative energy loss are subject to several limitations which restrict their applicability to more general situations. On the experimental side, these measurements at RHIC suffer from a trigger bias --- due to the requirement of a high transverse momentum particle in the event --- which affects the production mechanism. This bias is very difficult to consider in analytical models as it comes strongly related to energy-momentum constraints, while most models have been developed within high-energy approximations.

On the purely theoretical side, the existing developments have been mainly designed for dealing with one-particle inclusive distributions, with multi-gluon emission treated by simple ansatzs \cite{Baier:2001yt}. Although this prescription is reasonable for the inclusive quantities characterizing RHIC measurements until recently, it clearly lacks of the needed degree of refinement for the more general situations that we are facing. Additionally, although radiative energy loss still provides the most successful phenomenology of jet quenching, other effects, as collisional energy loss (e.g. \cite{Mustafa:2004dr}) could also play a role --- in fact, collisional energy loss would correspond to non-eikonal (finite energy) corrections to the computed gluon radiation spectra. Also, several implementations of radiative energy loss computed, in particular, using different prescriptions to treat the medium averages exist \cite{Majumder:2007iu}.

To better characterize the produced medium and to distinguish among these different possibilities, both more differential observables at large transverse momentum and unbiased measurements like jets, are required \cite{Salgado:2003rv}. While RHIC is starting to look at some of these new opportunities \cite{jets}, the Large Hadron Collider \cite{lhc} will be the ideal place due both to the higher collision energy and to the characteristics of the detectors.
 
 Radiative energy loss implies a modification of the standard QCD radiation pattern, whose generic expectations are a decrease of the energy of the leading particle in the jet (jet quenching), an enhancement of the intra-jet multiplicity and an increase of the typical emission angle with respect to the jet axis (jet broadening) \cite{rel,Salgado:2003rv}. While analytical attempts exist to study such modification and expectations within several approximation to the QCD cascading process (e.g. \cite{BorghiniWiedemann,Dremin:2006da}), the proper tool for considering the QCD branching process in the final state, with full energy-momentum conservation, is a Monte Carlo simulator.
 
 In spite of the fact that a
 probabilistic interpretation of radiation in a medium requires phenomenological
 assumptions, the practical advantages of a Monte Carlo are numerous. First,  
 it allows the access to observables other than the limited single
 inclusive measurements, such as different jet
 shapes, jet multiplicities, multiparticle intra-jet correlations, etc. Moreover, such an implementation makes it possible to explore new physical
 mechanisms in jet development, as the interplay of the multi-gluon
 radiation with the medium length, effects of the color flow and
 reconnections, effects of recoil with the medium and others. 
 A Monte Carlo tool is also of utmost importance for the experimental analysis. Codes as PYTHIA \cite{Sjostrand:2006za} or HERWIG \cite{Corcella:2000bw} are in the core of the software that the different experimental collaborations apply for correcting the data, e.g. in the calibration of the jet energy. Clearly, under these conditions, a code including medium effects in the jet development provides an essential contribution to the corresponding analyses in the heavy-ion program. Several implementations of radiative energy loss in Monte Carlo codes exist \cite{mc}. Each of them focus on specific aspects and relies on different simplifications.

 In this paper we present a Monte Carlo with medium-modified final-state radiation, based on the
 ideas described in \cite{msf} (a preliminary presentation of the Monte Carlo can be found in \cite{Armesto:2008qh}). There, medium effects enter as an additive correction to the standard, vacuum splitting functions. The modification is taken from the full medium-induced gluon radiation spectrum computed in the multiple-soft scattering approximation (also known as Baier-Dokshitzer-Mueller-Peign\'e-Schiff, BDMPS) and used in previous phenomenology of jet quenching \cite{rel}. In this case, the spectra depends on two quantities, which, for a static scenario are the length, $L$, of the medium and the transport coefficient, $\hat q$. The second quantity encodes all the information about the medium properties and is taken as a fitting parameter in practical applications. It also relates, through a single parameter, the inelastic energy loss and the angular broadening. This medium-modification is implemented at the level of the splitting functions in the standard final state showering routine {\tt PYSHOW} in PYTHIA.

A medium-modified parton shower should also include the interplay between the finite size of the medium and the space-time evolution of the jet. This interplay is known to play an important role to suppress radiation at the inclusive one-gluon level due to destructive interference from different scattering centers. The corresponding case for multi-gluon radiation, in particular the role of the different ordering variables, has not yet been solved from first principles. Here, we will adopt a hybrid approach in which we consider virtuality as a good ordering variable, but correct for the finite formation time of the gluons at each step during evolution. In its present form, our implementation does not consider effects as the recoil of the scattering centers, and consequently elastic energy loss is not taken into account, or the modification of the color structure of the cascade by exchanges with the medium. These, and other mechanisms, can be implemented by further modification of the shower routine {\tt PYSHOW} which we plan to do in the future.

 The paper is organized as follows:
In the next Section, we focus on the description of the implementation of medium-induced gluon radiation in the $t$-ordered final-state-radiation routine in PYTHIA. For a discussion of the basis of the modification of the splitting functions, we refer the reader to \cite{msf}. Some results are then presented in Section \ref{results}. We end with some conclusions and outlook in Section \ref{conclu}. Besides, some practical instructions to run the code are given in Appendix \ref{manual}.
A publicly available version of our routine, called Q-PYTHIA, can be found in \cite{pweb}.

\section{The Monte Carlo}
\label{montecarlo}

\subsection{Medium-modified splittings}

In this Subsection we briefly recall the treatment of the medium modification of parton splitting.
As discussed in detail in \cite{msf}, the spectrum for medium-induced gluon radiation \cite{rel} results in the sum of a vacuum term plus a medium-induced term:
\begin{equation}
\frac{dI^{\rm tot}}{dz\, d{\bf p}_T^2}=\frac{dI^{\rm vac}}{dz\, d{\bf p}_T^2}+\frac{dI^{\rm med}}{dz\, d{\bf p}_T^2}\,.
\label{sum}
\end{equation}
The former can be related with the small-$x$ limit of the standard Dokshitzer-Gribov-Lipatov-Altarelli-Parisi (DGLAP) splitting functions $P_{\rm vac}(z)$ \cite{dglap},
\begin{equation}
\frac{dI^{\rm vac}}{dz\, d{\bf p}_T^2}=\frac{\alpha_s}{2 \pi}
\frac{1}{{\bf p}_T^2} P_{\rm vac}(z),\ \ P_{\rm vac}(z)
\simeq \frac{2 C_R}{1-z}\,.
\label{vacsplit}
\end{equation}
In this two Eqs., $x=1-z$ is the momentum fraction of the parent parton  taken by the emitted gluon, ${\bf p}_T$ the transverse momentum of the emitted gluon with respect to the direction of its parent, and $C_R$ the quadratic Casimir of the color representation of the parent parton.

The modification of the DGLAP splitting function is now performed by matching the medium and the vacuum cases, Eqs. (\ref{sum}) and (\ref{vacsplit}), reading 
 \begin{equation}
P_{\rm tot} (z)= P_{\rm vac} (z)\to
 P_{\rm tot} (z)=P_{\rm vac}(z)+\Delta P(z,t,\hat{q},L,E),\ \ \Delta P(z,t,\hat{q},L,E)\simeq \frac{2 \pi  t}{\alpha_s}\, 
\frac{dI^{\rm med}}{dzdt}\, .
 \label{eq1}
 \end{equation}
 The correction $\Delta P(z,t,\hat{q},L,E)$ depends not only on the energy fraction
 $z$ but also on the virtuality $t=z(1-z){\bf p}_T^2$ of the radiating parton and its energy $E$, and on the medium characteristics relevant for radiative energy loss: transport coefficient $\hat{q}$ (which corresponds to the mean squared transverse momentum transferred from the medium to the parton traversing it per mean free path) and  medium length $L$. For practical reasons, we will use as variables characterizing the medium the accumulated transverse momentum $\hat{q}L$ and the gluon characteristic frequency $\omega_c=\hat{q}L^2/2$. The medium-induced gluon radiation spectrum $dI^{\rm med}/{dz\, d{\bf p}_T^2}$ will be taken in the multiple-soft scattering (or BDMPS) approximation \cite{rel}. 
 
 This ansatz constrains only the small-$x$ part of the correction $\Delta P(z,t,\hat{q},L,E)$. Different large-$x$ extensions have been essayed, yielding similar results as discussed in \cite{msf}. We will use the default extension discussed there, consisting in multiplying the medium-induced gluon spectrum by $(1+z^2)/2$ if the parent parton is a quark and by $z$ if the parent parton is a gluon (in this case we also impose a symmetrization around $z=1/2$).
 
The additivity of a vacuum and a medium contribution is a common feature of the spectra when computed for the one-gluon inclusive case. Medium-modified splitting functions in a higher-twist formalism have been found to present the same factorization under the assumption of exponentiation of the elementary probabilities \cite{Wang:2001if}. Nevertheless, a formal proof or disproof of this separation and of the factorization of no-splitting and splitting probabilities (respectively given by the Sudakov form factor, see below, and by the DGLAP splitting functions), like those available in the vacuum
--- see e.g.  \cite{Konishi:1979cb} and references therein ---, is still missing for a radiating parton traveling through a colored medium.
 
\subsection{Basics steps for a Monte Carlo implementation}

Basically, a branching algorithm must solve the following problem: given a
parton coming from a branching (or production) point with coordinates ($t_{1}$,$x_{1}$), with  $t_{1}$ the virtuality and
$x_{1}$ its energy fraction, which are the coordinates ($t_{2}$,$x_{2}$) for the next branching? 

Ignoring, for simplicity, parton labels, the Sudakov form factor 
\begin{equation}
\Delta(t_{1})=\exp{\left[-\int_{t_0}^{t_{1}} {dt^\prime \over t^\prime}
\int_{z^{-}}^{z^{+}} dz {\alpha_s(t_{1})
\over 2 \pi} P(z)\right]}
\label{eq2}
\end{equation}
 gives the probability for a parton not
to branch (in a resolvable manner) while evolving from an initial scale $t_{0}$ to another scale
$t_{1}$. Consequently, $\Delta(t_{2})/\Delta(t_{1})$ stands for the
probability of evolving from $t_{1}$ to $t_{2}$ without branching. Thus
$t_{2}$ can be generated by solving the equation
\begin{equation}
 \frac{\Delta(t_{2})}{\Delta(t_{1})}=R ,
 \label{eq3}
 \end{equation}
$R$ being a random number between 0 and 1.

The energy fraction kept by the parton in the next branching $z_{2}$ can be
diced down by solving the equation:
\begin{equation}
\int_{z_{-}}^{z_{2}} dz \frac{\alpha_{S}}{2\pi} P(z)=R^\prime
\int_{z_{-}}^{z_{+}} dz \frac{\alpha_{S}}{2\pi} P(z), \label{eq4}
\end{equation}
with $R^\prime$ another random number between 0 and 1. Eqs.
\eqref{eq3} and \eqref{eq4}  are the two basic steps of a branching Monte Carlo algorithm. 
 
Now let us address in more detail the procedure followed by PYTHIA. 
The scale $t$ of the evolution in PYTHIA is chosen to be the invariant mass
squared of the decaying particles. 
The showering is performed by the creation of pairs of partons, conserving
four-momentum at each step. 
If, in the course of the evolution, a parton $a$
reaches a scale $t_{a}$ that is smaller than any minimal mass of any allowed
branching mode, then its mass and its scale will be set to its current mass
$m_{a}^{0}$ and it will leave the parton shower without further branching.
The evolution has a scale cut-off $t_{0}=Q_{0}^{2}$.

In this setup, it is now conceptually simple to introduce our medium-modified splitting
functions \eqref{eq1} which enter at steps \eqref{eq3} and \eqref{eq4}  described above.
A straightforward implementation is, however, not possible because of technical reasons: PYTHIA approximates the splitting functions by their $z\to 1$ form to simplify the dicing procedure; this approximation is later corrected by a rejection method. The first change we make is then to solve all these integrals numerically at every step. So, this procedure modifies standard PYTHIA even for the vacuum case. In Fig. \ref{fig:1} we check that our modification of PYTHIA, in which we implement the exact vacuum splitting function without the $z\to 1$ approximation, matches very well the results of usual PYTHIA. Once this modification is performed, we add the medium term to the vacuum splitting functions at every step of the evolution. Note also that our medium modifications are for $g\to gg$ and $q(\bar q)\to q(\bar q)g$
splittings \cite{msf}. We do not consider any modification of the  $g\to q\bar{q}$ branching because
its splitting probability is not singular at $z\to 1$. 

In Eqs. \eqref{eq2} and \eqref{eq4} the value of the maximum initial virtuality, $t_{max}\propto E_{jet}
^2$, of the lower virtuality limit, $t_0=1$ GeV$^2$, the lower and upper limits of the $z$-integrals, $z_\pm\equiv z_\pm(t)$, and the scale at which $\alpha_s$ runs, as well as all other aspects of the $t$-ordered evolution (for example, the implementation of angular ordering which is a default), are the PYTHIAv6.4.18 defaults, see \cite{Sjostrand:2006za}. They can be changed by the user. 

\FIGURE{
  \includegraphics[width=13.2cm]{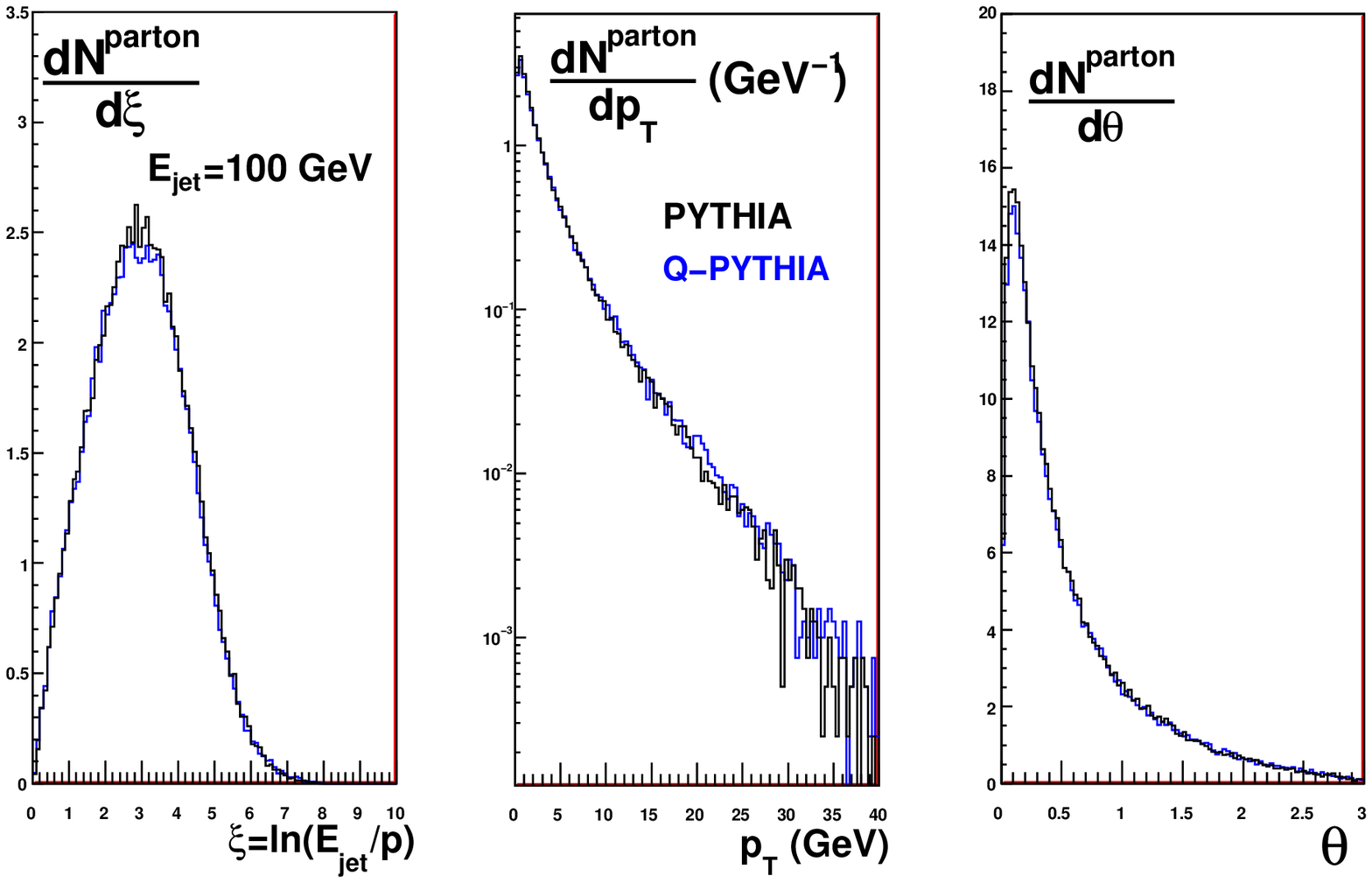}
\caption{Intra-jet parton distributions in $\xi=\ln{(E_{jet}/p)}$ (left), $p_{T}$ (middle) and $\theta={\rm acos}(p_z/p)$ (right)
   for a gluon of initial energy $E_{jet}=100$ GeV. PYTHIA default (black lines)  and
   our results with $\hat{q}=0$ (blue lines) are compared.}
\label{fig:1}       
}

Finally, let us note that  we have implemented the medium modifications only in the $t$-ordered routine for final state radiation. Actually, three variables have be chosen in the literature as evolution variables: virtuality $t$, squared transverse momentum ${\bf p}_T^2$ and angle $\theta$. The fortran implementation PYTHIAv6.4 sets $t$ as the default evolution variable, while it offers the possibility of using ${\bf p}_T^2$ which has become the default in the C++ version PYTHIAv8 \cite{Sjostrand:2006za}. On the other hand, angle $\theta$ is used in HERWIG++ \cite{Bahr:2008pv} and $t$ (plus angular ordering) in SHERPA \cite{Gleisberg:2008ta}. 
Our medium modification of the splitting functions, Eq. (\ref{eq1}), is completely general and applicable to any evolution variable. We plan its implementation in different variables in future works \cite{accs}.

\subsection{Length and energy evolution}
\label{seclen}

The presence of a medium makes it necessary to consider the interplay between the momentum evolution variable (virtuality in our case) and the space-time dimensions of the medium. For single gluon emission the presence of a finite-size medium is known \cite{rel} to produce an interference effect known as the Landau-Pomeranchuk-Migdal effect. A full treatment of such problem for multi-gluon emission is still unsolved theoretically and, in fact, it is linked to the factorization problem mentioned previously.
Here, we will use an assumption which relies on the factorization underlying our approach: we will treat the ordering as in usual PYTHIA, but introducing  a correction procedure at every splitting to take into account the space-time evolution of the shower. We consider, as well, the energy degradation of the shower at each splitting --- notice that the medium-modified splitting functions (\ref{eq1}) depend now on the energy of the parent parton.  

More specifically, the length traveled by a parton before a gluon decoheres from its wave
function and is radiated, can be estimated \cite{rel} by the
gluon formation length $l_{coh}={2 \omega_i/{\bf k}_{T,i}^2}$, where $\omega_i$ and
${\bf k}_{T,i}$ are the energy and transverse momentum (with respect to its parent parton)
of the $i$-th emitted gluon,
respectively.   

The shower begins with a parton that faces the full length of the medium $L$, so
the medium effects on the probability of the first branching are evaluated at $L$. The coherence
length of the emitted gluon is then computed being its next branching
evaluated at $L-l_{coh}$. The process is iterated. Possible effects of dynamical expansion of the medium are handled analogously by considering that gluons are formed at times given by their formation lengths.

Also the energy degradation is considered. For a process $a(E_b+E_c)\to b(E_b)+c(E_c)$, the medium effects in the branching process of $a$ is considered at energy $E_b+E_c$, while the subsequent branchings of $b$ and $c$, if any, are considered at $E_b$ and $E_c$ respectively. In our default results for the medium, both the evolution in length and the energy degradation are considered. The separate effect of these aspects of evolution will be discussed below.

All these modifications have been implemented in the $t$-ordered final state radiation routine {\tt PYSHOW}, available as a code named Q-PYTHIA in \cite{pweb}. A brief manual is provided in Appendix \ref{manual}. Note that neither the modification of the medium length nor the energy degradation are considered in usual analytical approaches to radiative energy loss \cite{rel,Baier:2001yt,msf}: all splittings are considered for a parton with the same energy and medium length or density.

\section{Results}
\label{results}

As mentioned in the Introduction, the generic expectations of medium-induced gluon radiation formalisms are:
\begin{itemize}
  \item A softening of the spectra: jet quenching.
  \item An increase of the multiplicity. 
  \item An angular broadening of the jet: jet broadening.
\end{itemize}
These are the main features of parton energy loss that should be reflected in our
implementation. However, the energy-momentum conservation within the parton shower may play a distorting role on these naive expectations based on the one-gluon inclusive spectra, as we will see later. Notice, also, that an energy-momentum flow between the parton shower and the medium (not taken into account in our present implementation) could further affect the cascade, especially for the softer particles.

To see the results, we run PYTHIA with our medium modifications.
For this model description, we first focus on representative results which illustrate the effects of medium-induced gluon radiation in the branching process, to provide in the last Subsection some examples of practical interest. The statistics we have used is $10^5$ generated events for each case. Unless otherwise indicated, the maximum initial virtuality has been taken as $t_{max}=2E^2$, with $E$ the energy of the parton which initiates the shower.

\subsection{Results at parton level}
\label{partons}

To see the results at parton level, we run our medium-modified {\tt PYSHOW} on a gluon of
energy $E_{jet}=10$ and 100 GeV moving along the positive $z$-axis, and we study the intra-jet distribution of
final partons in energy fraction (actually in $\xi=\ln{(E_{jet}/p)}$ with $p=|\vec{p}|$ the modulus of the momentum of the final particle --- the hump-backed plateau plot), transverse momentum $p_T=\sqrt{p_x^2+p_y^2}$ of the final particle, and polar angle $\theta={\rm acos}(p_z/p)$ of the final particle. The medium length is fixed to $L=2$ and 5 fm, considering an homogeneous piece of matter defined by the planes $(x,y,0)$ and $(x,y,L)$. 

\FIGURE{
  \includegraphics[width=13.2cm]{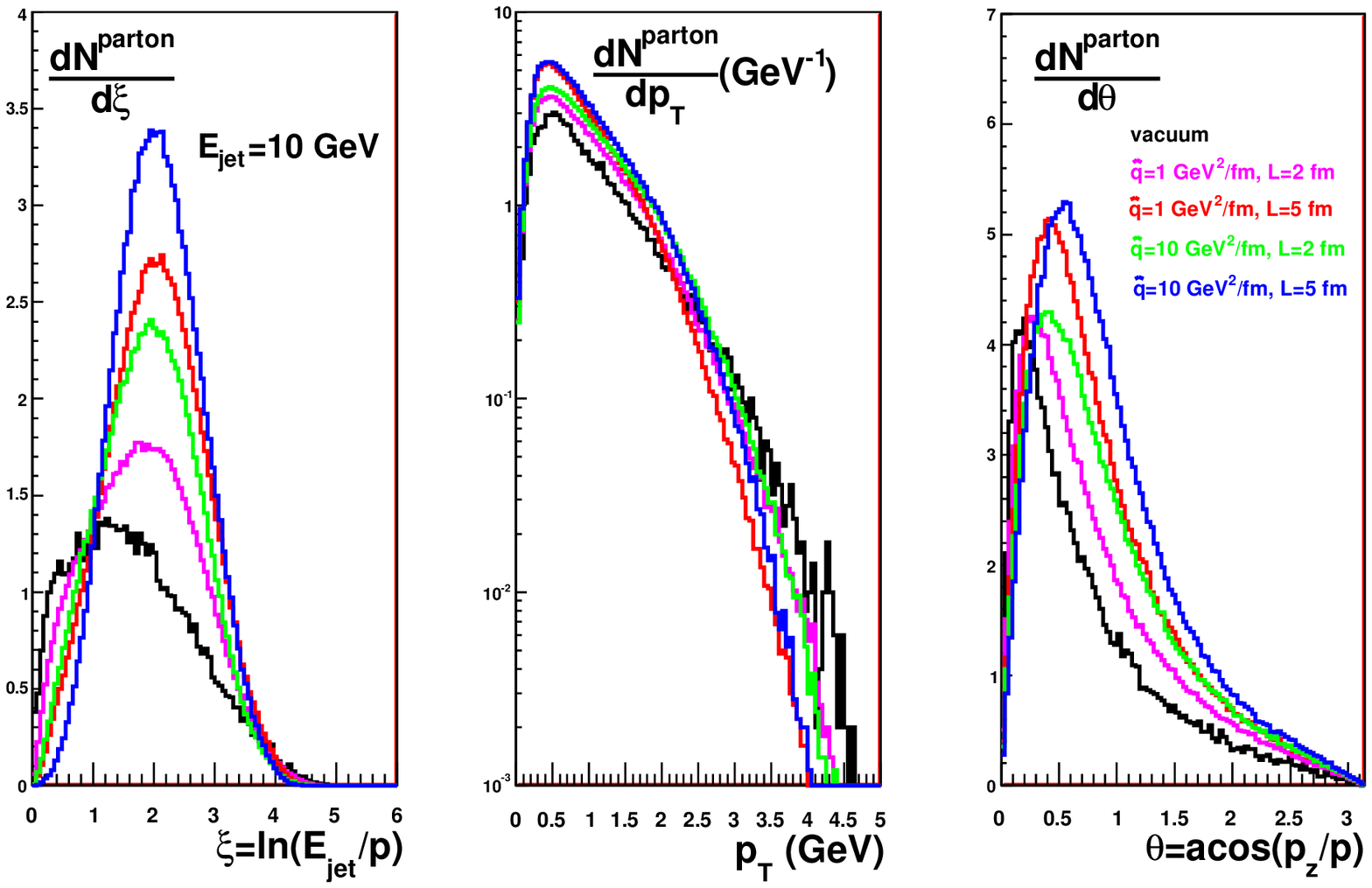}
      \includegraphics[width=13.2cm]{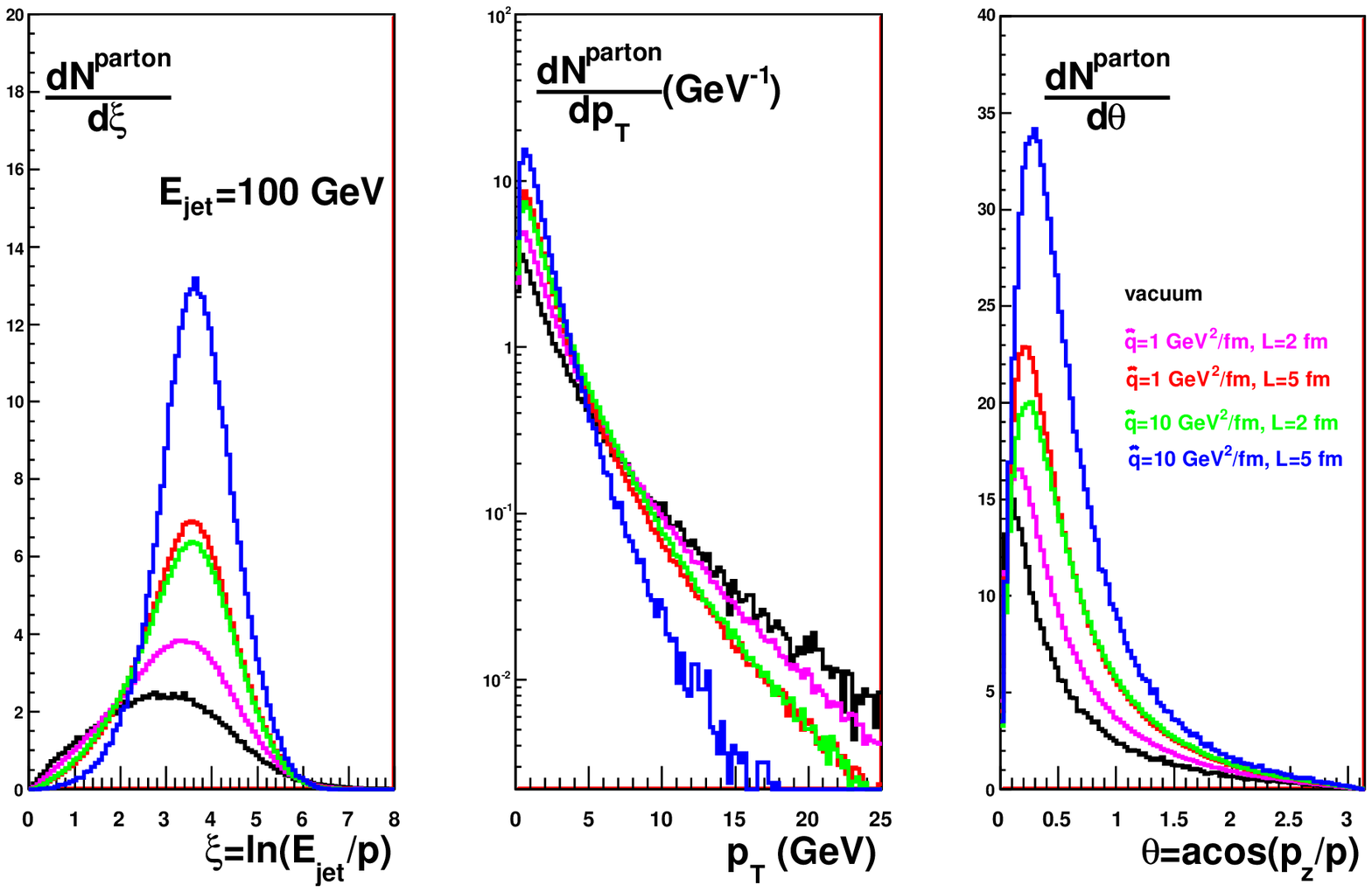}
\caption{Upper plots: Intra-jet parton distributions in $\xi$ (left), $p_{T}$ (middle)
 and $\theta$ (right) for a gluon of initial energy $E_{jet}=10$ GeV in a medium of
 length  $L=2$  and 5 fm and for different transport coefficients $\hat{q}=0$, 1 and 10 GeV$^2$/fm, see the legend on the plot. Lower plots:  Id. but for $E_{jet}=100$ GeV.}
\label{fig:2}      
}

The results are shown in Fig.  \ref{fig:2}. We observe a suppression of low-$\xi$ particles and a large
enhancement of particles with large $\xi$-values, as expected due to the energy degradation of the leading particle (energy loss).
We also observe a suppression of high-$p_{T}$
particles and the corresponding enhancement of intermediate-$p_{T}$ particles.
The $p_T$-spectrum should be softer than vacuum at low transverse momentum since low-$p_{T}$ particles should be  kicked towards
higher values of the transverse momentum. However we find a clear enhancement
of low-$p_{T}$ particles. Here the energy-momentum conservation in PYTHIA is causing a large effect as any increase of the multiplicity means that the total jet energy is shared among a larger number of partons. The angular distribution broadens also with increasing transport coefficient, as expected.

\subsection{Effects of hadronization}
\label{hadrons}

\FIGURE{
  \includegraphics[width=12.8cm]{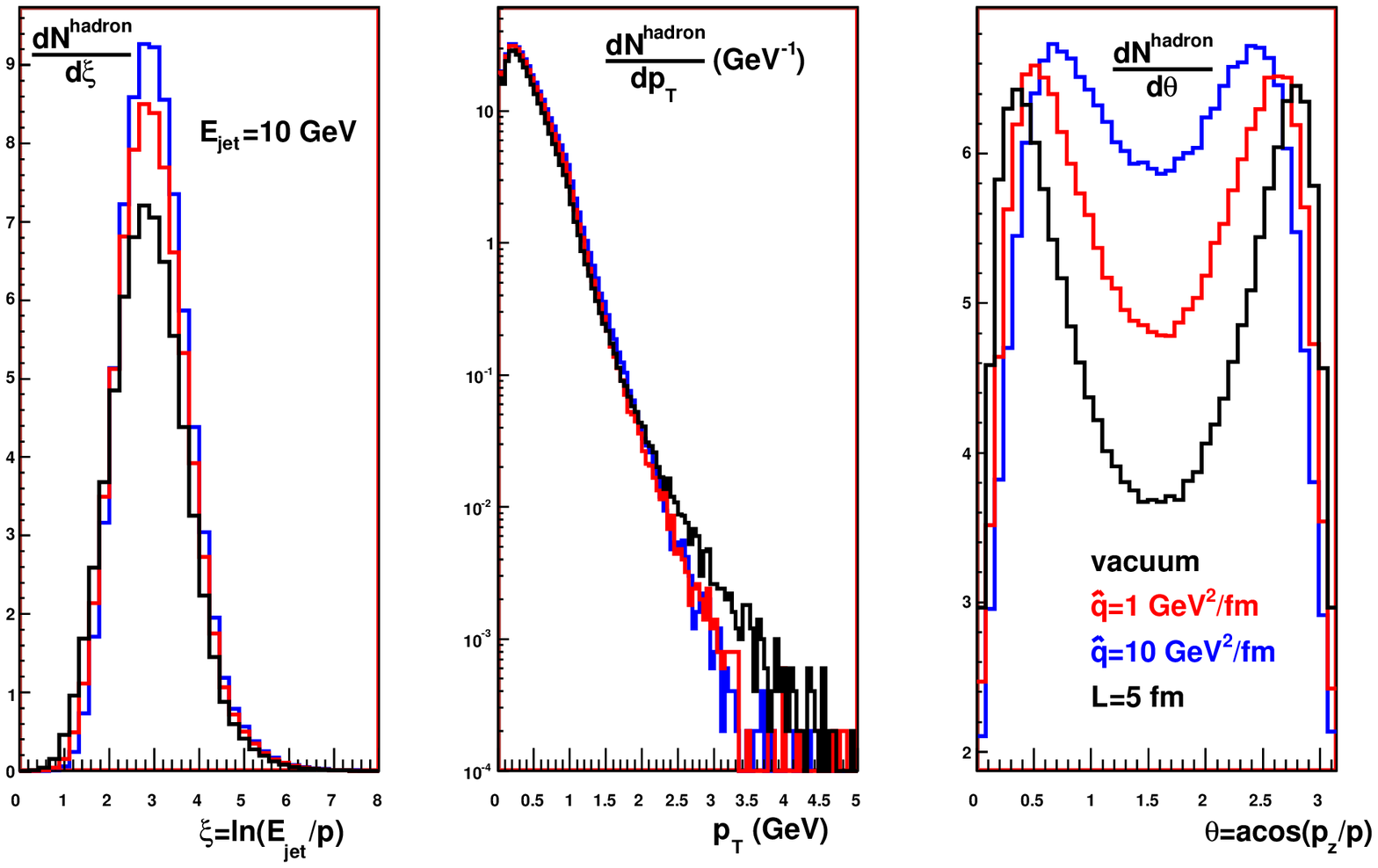}
   \includegraphics[width=12.8cm]{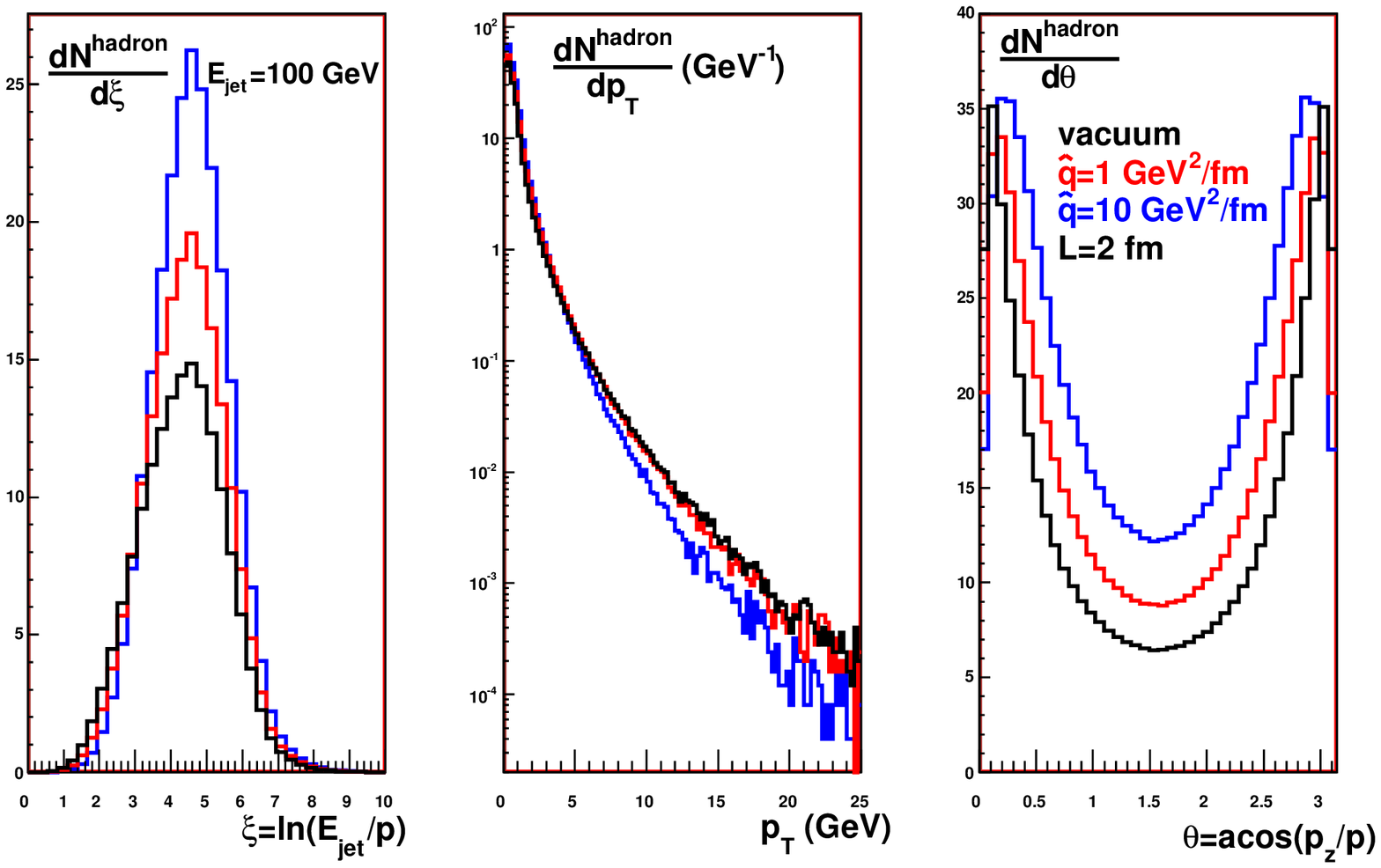}
\caption{Upper plots: Intra-jet hadron distributions in $\xi$ (left), $p_{T}$ (middle)
 and $\theta$ (right) for two back-to-back gluons of initial energy $E_{jet}=10$ GeV, see the text, in a medium of
 length  $L=5$ fm and for different transport coefficients $\hat{q}=0$ (black), 1 (red) and 10 (blue lines) GeV$^2$/fm. Lower plots: Id. for $E_{jet}=100$ GeV and $L=2$ fm.}
\label{fig:3}       
}

To introduce
hadronization effects, two back-to-back gluons of the same energy $E_{jet}$ running along the $z$-axis with opposite directions, in color singlet, are considered.
PYTHIA hadronization routines are run after final state radiation. All final hadrons as produced by standard PYTHIA are taken into account. Now we consider an homogeneous piece of matter defined by the planes $(x,y,-L)$ and $(x,y,L)$, with the back-to-back pair initially produced at $(0,0,0)$.

The corresponding results are shown in Fig. \ref{fig:3}. The multiplicity enhancement at large $\xi$ and small $p_T$ is less pronounced, particularly at small energies.
We conclude that, as expected, hadronization tends to wash out medium effects in the soft region.
The broadening of the angular distribution remains, and the region perpendicular to the initial parton directions is partially filled.

\FIGURE{
    \includegraphics[width=12.5cm]{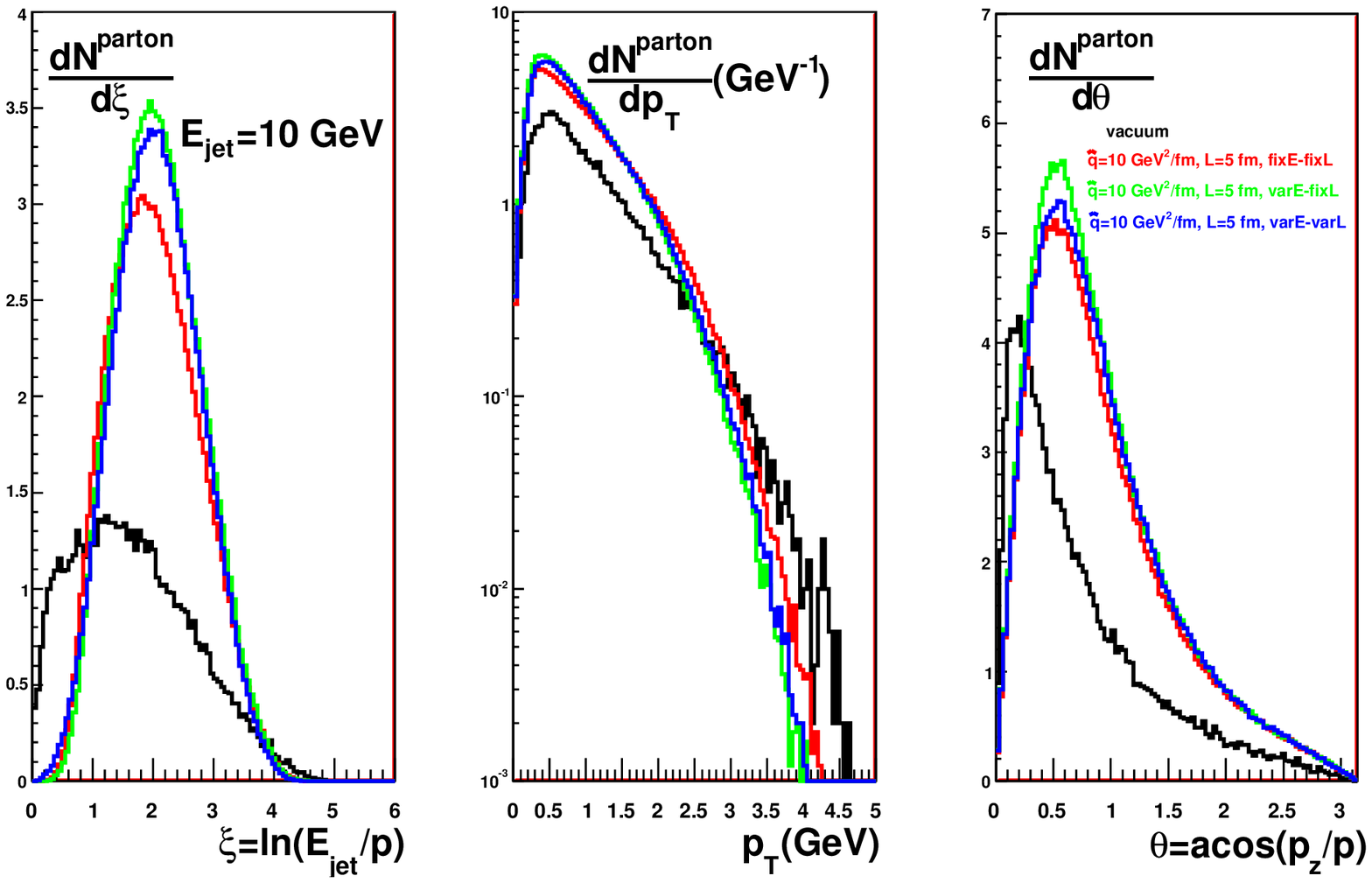}
        \includegraphics[width=12.5cm]{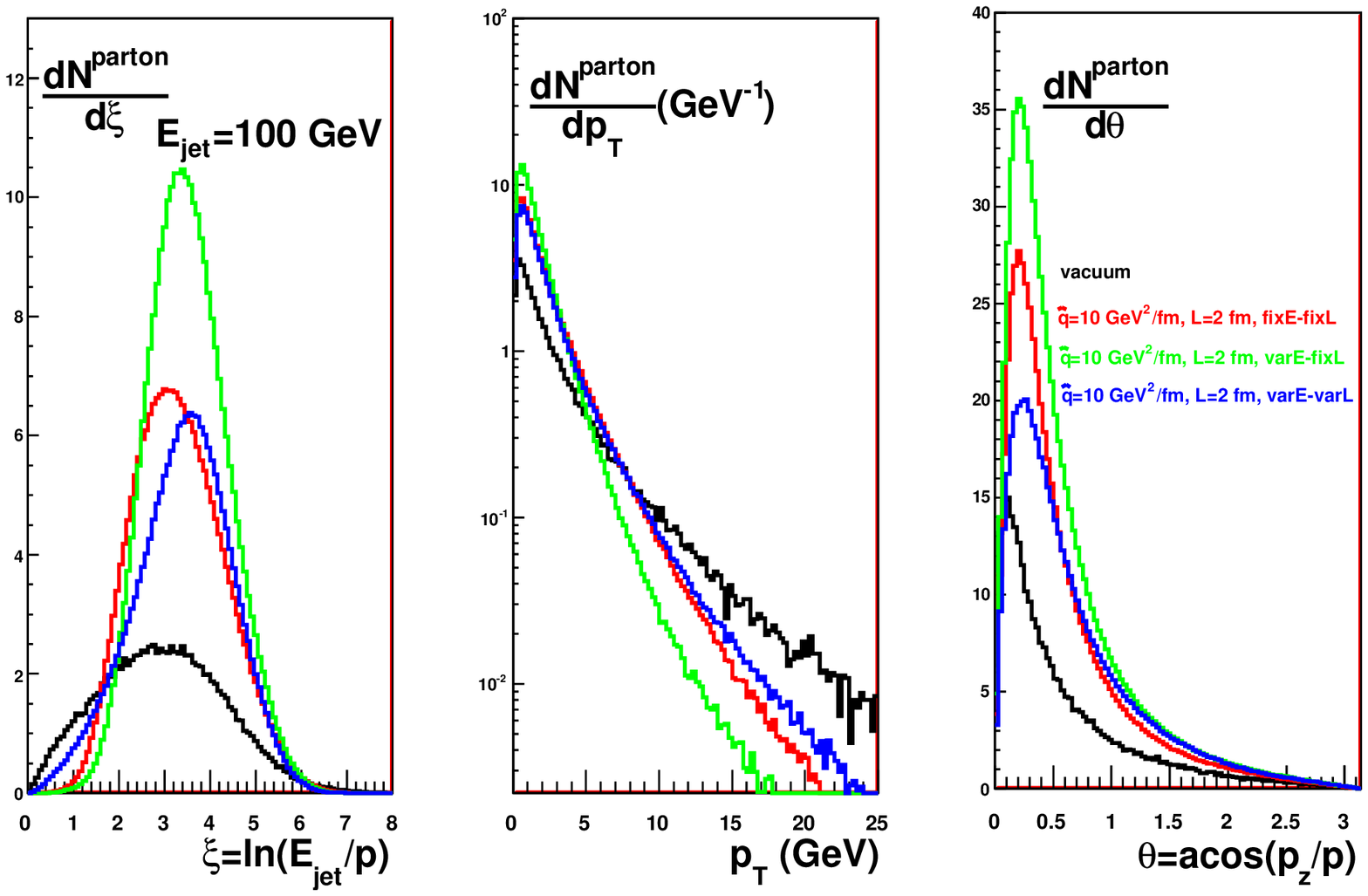}
\caption{Upper plots: Intra-jet parton distributions in $\xi$ (left), $p_{T}$ (middle)
 and $\theta$ (right) for a gluon of initial energy $E_{jet}=10$ GeV in a medium of
 length  $L=5$ fm and for different transport coefficients $\hat{q}=0$ (black) and 10 (colored lines) GeV$^2$/fm. In
  red we show the results for no evolution in energy or length (1), in green those for evolution in energy but not in length (2), and in blue those for evolution in energy and length (3), see the text. Lower plots: Id.  but for $E_{jet}=100$ GeV and  $L=2$ fm.}
\label{fig:4}       
}

\subsection{Different effects on evolution: medium length and energy degradation}

In Fig. \ref{fig:4} we study the different effects discussed in Subsection \ref{seclen}. More concretely, we show three cases:  (1) a shower with neither energy degradation  nor evolution in length, so medium effects are computed at the initial values in all branching ('no evolution in energy or length'); (2) considering the energy degradation but not the
evolution in length ('evolution in energy but not in length'); (3) our default: energy degradation and evolution in length ('evolution in energy and length'). 
We see that the medium effects are the largest in case (2). This is due to the
fact that radiative energy loss is roughly independent of the parton energy \cite{rel} so, as
the energy of the parton is reduced at each branching, they become more and
more important.  We also see that the evolution in length, case (3), causes a decrease
of the medium effects with respect to the previous case, as expected.

\subsection{Four examples: fragmentation functions, $R_{AA}$, jet shapes and jet multiplicities}

In Subsections \ref{partons} and \ref{hadrons} we have shown several results which illustrate the impact of our modification of the final-state parton shower on the hump-backed plateau, transverse momentum and angular distribution of partons and hadrons. In this Subsection we will indicate how some other observables are modified. We do not aim here to any quantitative comparison with present or future experimental situations, a work which is left for future studies.

First, we show in Fig. \ref{fig:5} the medium modification of the fragmentation function for $\pi^0$'s and hadrons expressed through the ratio of the fragmentation function in medium over that in vacuum. These fragmentation functions have been obtained through hadronization using the same setup as in Subsection \ref{hadrons}: two back-to-back gluons in color singlet. Actually the information is exactly the same as that contained in the hump-backed plateau, though plotted in different variables. The effects already  observed in the hump-backed plateau: enhancement of small-$z$  and suppression of high-$z$ hadrons with increasing $\hat q$ or $L$, are evident here.

\FIGURE{
    \includegraphics[width=9cm]{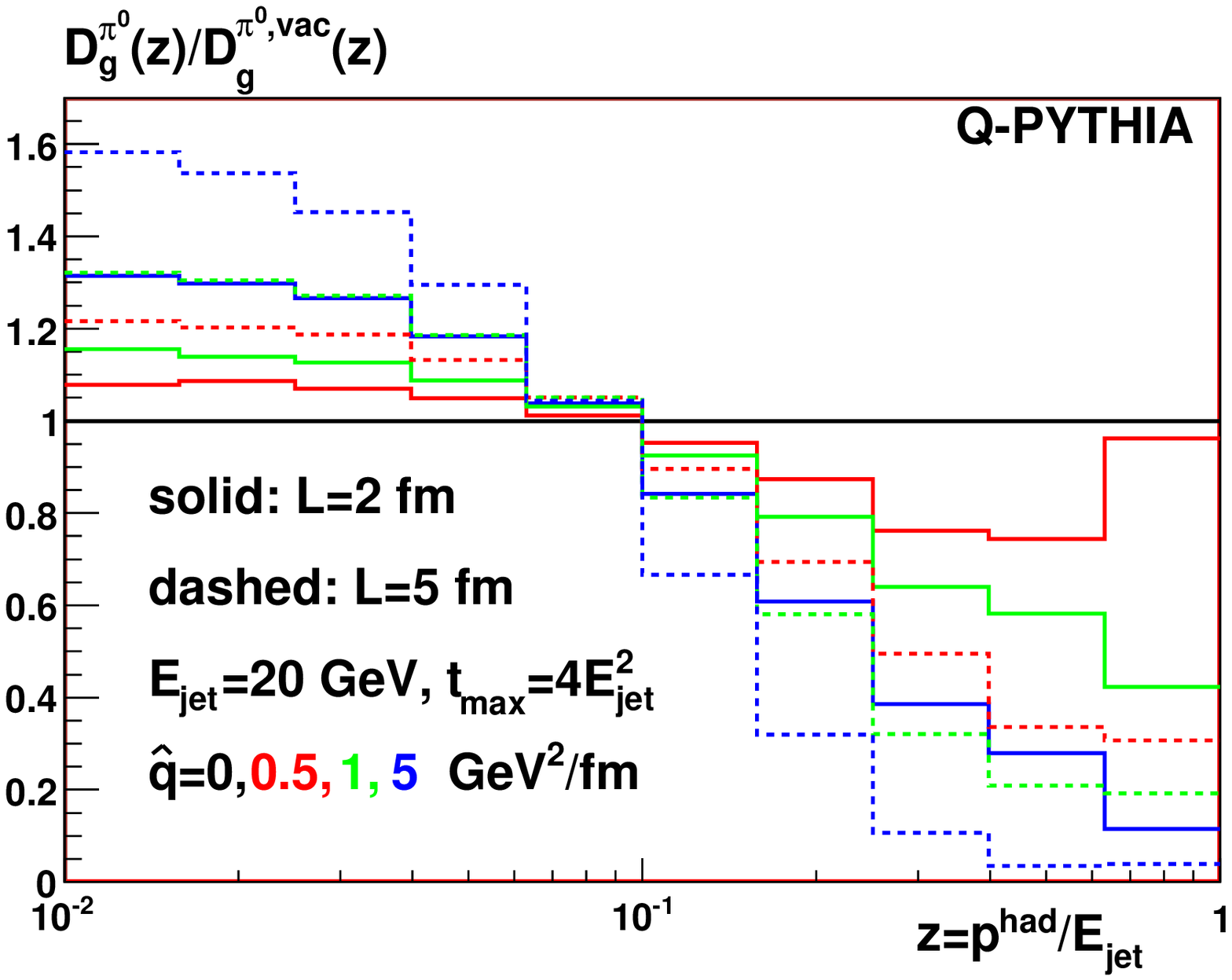}\includegraphics[width=9cm]{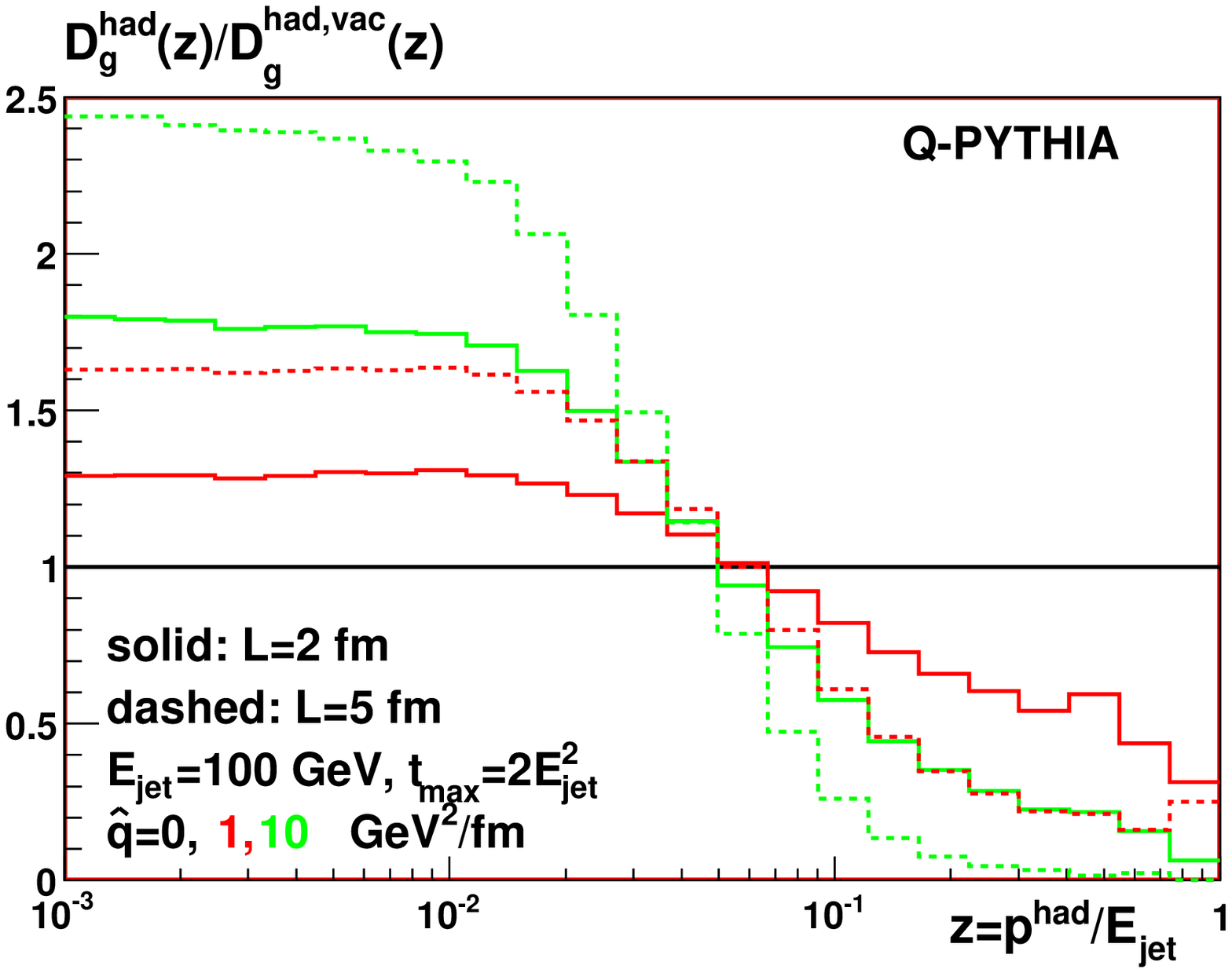}
\caption{Ratio of medium to vacuum fragmentation functions for $\pi^0$'s (left) and all hadrons (right), for different gluon energies $E_{jetÊ}$, medium lengths $L$, transport coefficients $\hat q$ and maximum virtualities $t_{max}$, see the legends on the plots.}
\label{fig:5}       
}

Second, we show results for the nuclear suppression factor, defined as the ratio between the number of particles measured over the expectation from a superposition of independent nucleon-nucleon collisions.
Its deviation from 1 is a reflection of the nuclear effects. The measurements at RHIC \cite{rhic} of its value being much smaller than 1 in central AuAu collisions, was the indication of jet quenching. In Fig. \ref{fig:6} we present results for the nuclear suppression factor for charged particles defined as
\begin{equation}
R_{AA}(\eta=0,p_T)=\frac{\left.\frac{dN^{ch}}{d\eta dp_{T}}{\rm (quenched)}\right|_{\eta=0}}{\left.\frac{dN^{ch}}{d\eta dp_{T}}{\rm (unquenched)}\right|_{\eta=0}\,}\, ,
\end{equation}
for which we run $10^6$ pp events at $\sqrt{s_{NN}}=200$ GeV both in the unquenched case ($\hat q=0$, $L=0$) and in the quenched case, requiring a minimum $p_T^{\rm min}=8$ GeV in PYTHIA\footnote{I.e. we consider pp events happening both in the vacuum and in a medium, where the only difference is the treatment of the final-state parton shower. In our implementation no modification of the underlying event has been included, nor even the superposition of different nucleon-nucleon collisions within the same event.}. For the latter the geometry is that of a $0-10 \%$ central PbPb 
collisions and the treatment of the production points and of the quenching parameters is done like in the PQM 
model \cite{Dainese:2004te}. In short, the production points of the hard scatterings are distributed 
in the nuclear overlapping area according to the probability of binary nucleon-nucleon collisions. Then, 
$\hat{q}$ and in-medium path length $L$ are computed locally through two integrals of the density of binary nucleon-nucleon collisions along parton trajectories isotropically distributed in azimuth, see the Appendix \ref{manual}. In this model the only free parameter is the 
scale of the transport coefficient $k$ (in fm). Using the value reported in Ref.  \cite{Dainese:2004te} to reproduce single-inclusive RHIC $R_{AA}$,  $k=6\cdot 10^6$ fm 
(which corresponds to an average $\langle\hat{q}\rangle=14$ GeV$^2$/fm in this case) 
we get a suppression factor of order 5 at $p_{T}>5$ GeV in 
semi-quantitative agreement with RHIC experimental data \cite{rhic}. 
%
\FIGURE{
    \includegraphics[width=12cm]{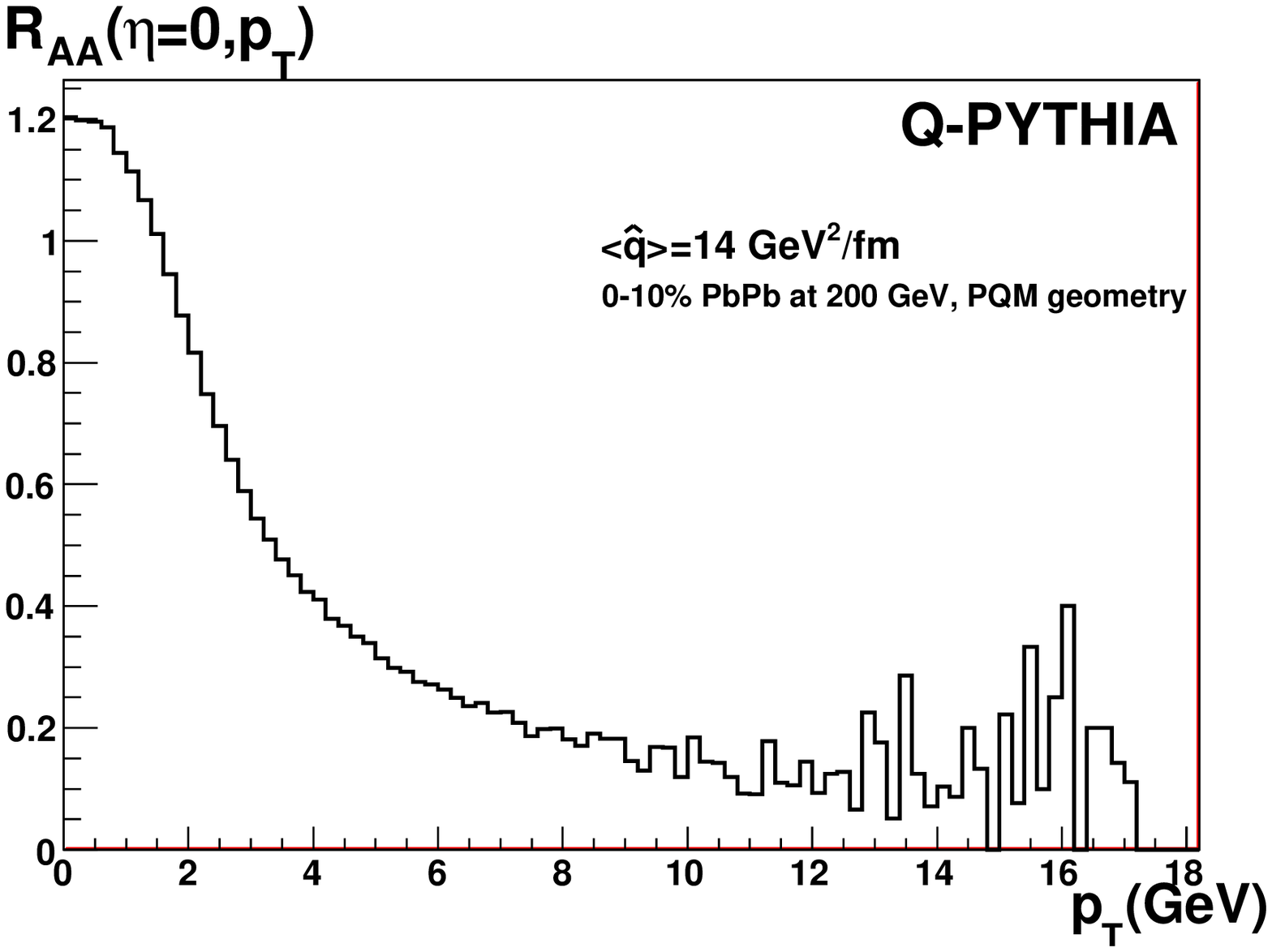}
\caption{Nuclear suppression factor $R_{AA}$ for charged hadrons  in $0-10 \%$ central PbPb
collisions at $\sqrt{s_{NN}}=200$ GeV. See the text for explanations.}
\label{fig:6}       
}
%
While no comparison to experimental data is aimed here, we note that these results are in good agreement with those from Ref. \cite{Dainese:2004te} in the PQM model which considers the energy loss of the leading parton through the simple ansatz \cite{Baier:2001yt} usually assumed in previous jet quenching phenomenology at RHIC. This implies that the introduction of evolution in virtuality and of energy-momentum conservation does not modify strongly, as also observed in \cite{msf}, the conclusions about medium characteristics at RHIC extracted in previously existing frameworks.

Third, we present some results on jet shapes. In Fig. \ref{fig:7} we compute the jet shape in pp collisions, with and without medium corrections, at $\sqrt{s_{NN}}=5.5$ TeV using the anti-$k_T$ recombination 
algorithm \cite{Cacciari:2008gp} within the FastJet package \cite{Cacciari:2005hq}.  For the produced particles considered for jet definition, we follow the Les Houches accord \cite{lha} (considering all decays within 10 mm from the production point). We proceed by finding the hardest jet 
in the event with radius $R^{j}=R^{max}=1.5$ and transverse momentum 
$p_{T}^{j}$. Its constituents are recursively reclustered with smaller resolution 
$R^{sj}<R^{j}$ into subjets of transverse momentum $p_{T}^{sj}$. The jet shape 
$\phi(r)=\langle p_{T}\rangle^{sj}/\langle p_{T}\rangle^{j}$ is the fraction of the jet transverse momentum inside the hardest 
subjet. For the generation of jets, at minimum $p_T^{min}= 100$ GeV is required in PYTHIA. Quenching is implemented for the geometry $0-10 \%$ central PbPb collisions as it was done for the previous example of the computation of $R_{AA}$.

\FIGURE{
    \includegraphics[width=12cm]{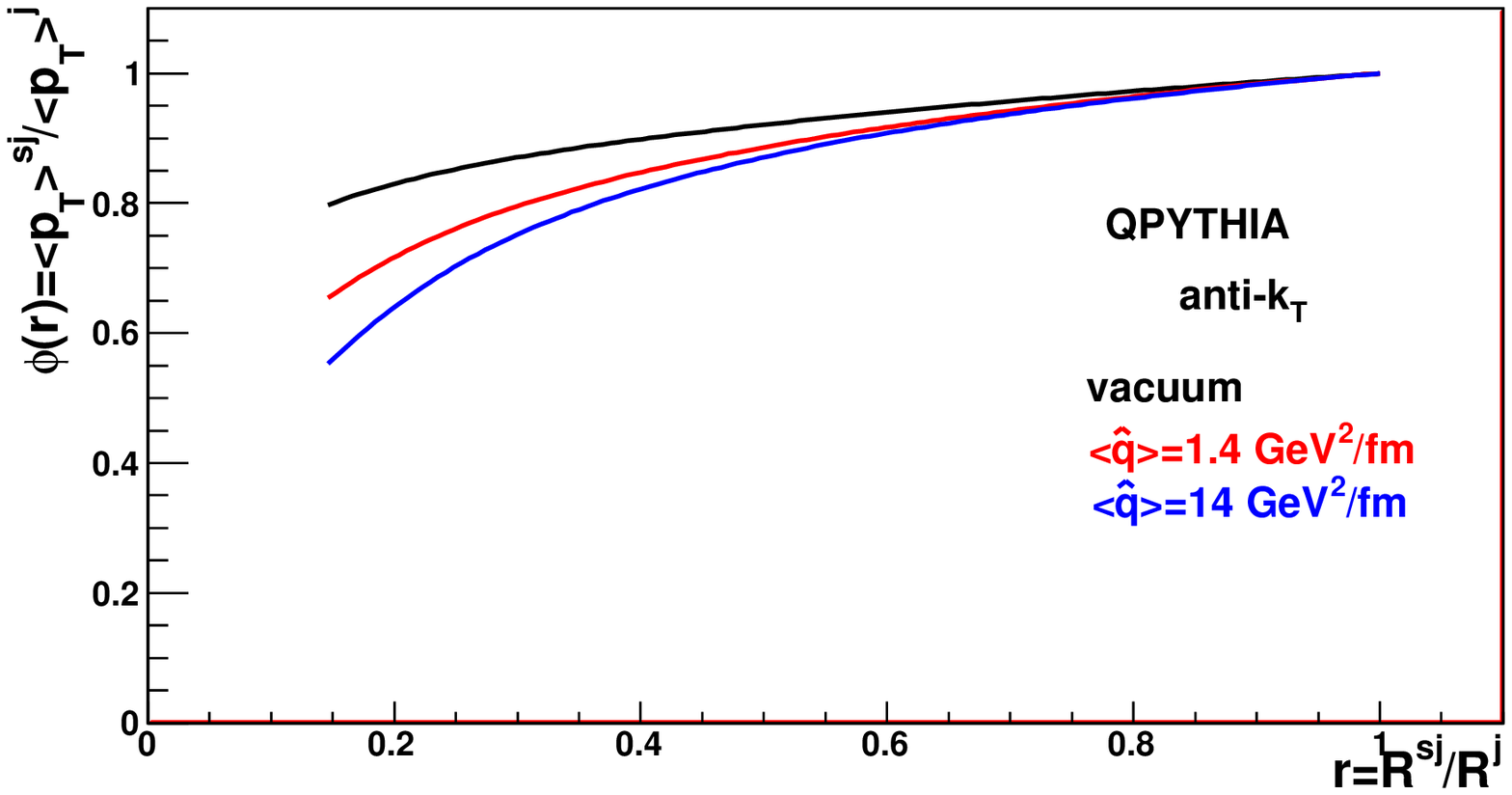}
\caption{Jet shapes in pp collisions at $\sqrt{s_{NN}}=5.5$ TeV using the anti-$k_T$ recombination 
algorithm for different quenching parameters, see the legend on the plot, and the same geometry as in Fig. \protect\ref{fig:6}. See the text for explanations.}
\label{fig:7}       
}

We find that in a vacuum jet, $\simeq 80 \%$ of the initial jet transverse energy is kept 
inside a subjet of radius $R^{sj}=0.3$ (corresponding to $r=R^{sj}/R^j=0.2$ in the figure). Quenching leads to a 
redistribution and broadening of the jet transverse energy: according to 
our plot,  $\simeq 60 \%$ of the initial jet energy is resolved 
with a subcone of radius $R^{sj}=0.3$ in the case of $\langle \hat q\rangle= 14$ GeV$^2$/fm.

Finally, we show in Fig. \ref{fig:8} the distributions in jet multiplicity (i.e. the number of reconstructed jets in an event) for the partonic decay products of a gluon jet of energy 100 GeV, using the anti-$k_T$ recombination 
algorithm, versus the radius of the algorithm. As expected, for fixed radius the distributions with a higher number of jets are enhanced by medium effects. Implementing cuts on the minimum $p_T$ of the reconstructed jets shifts all the distributions towards smaller radii.
subtraction
\FIGURE{
    \includegraphics[width=12cm]{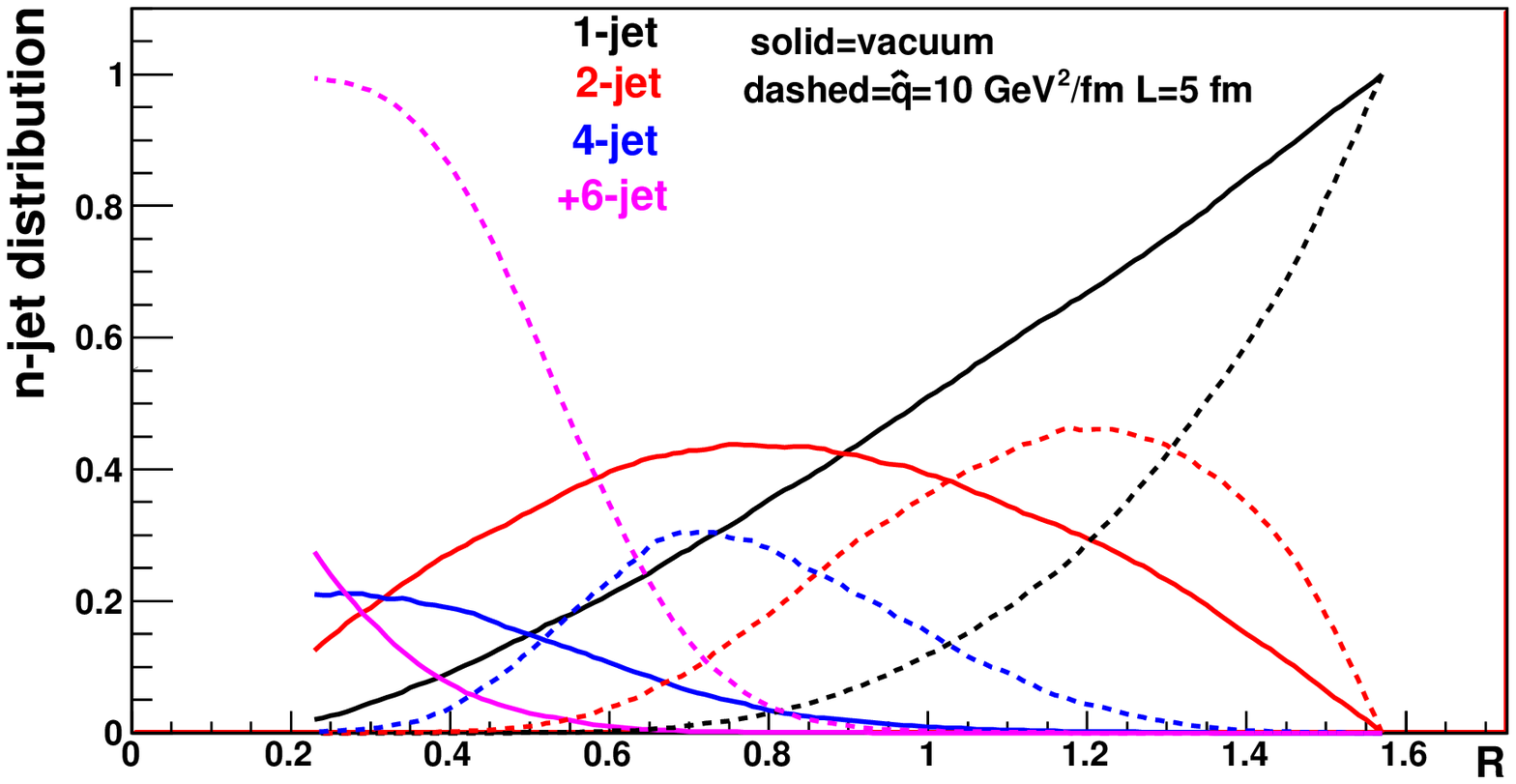}
     \includegraphics[width=12cm]{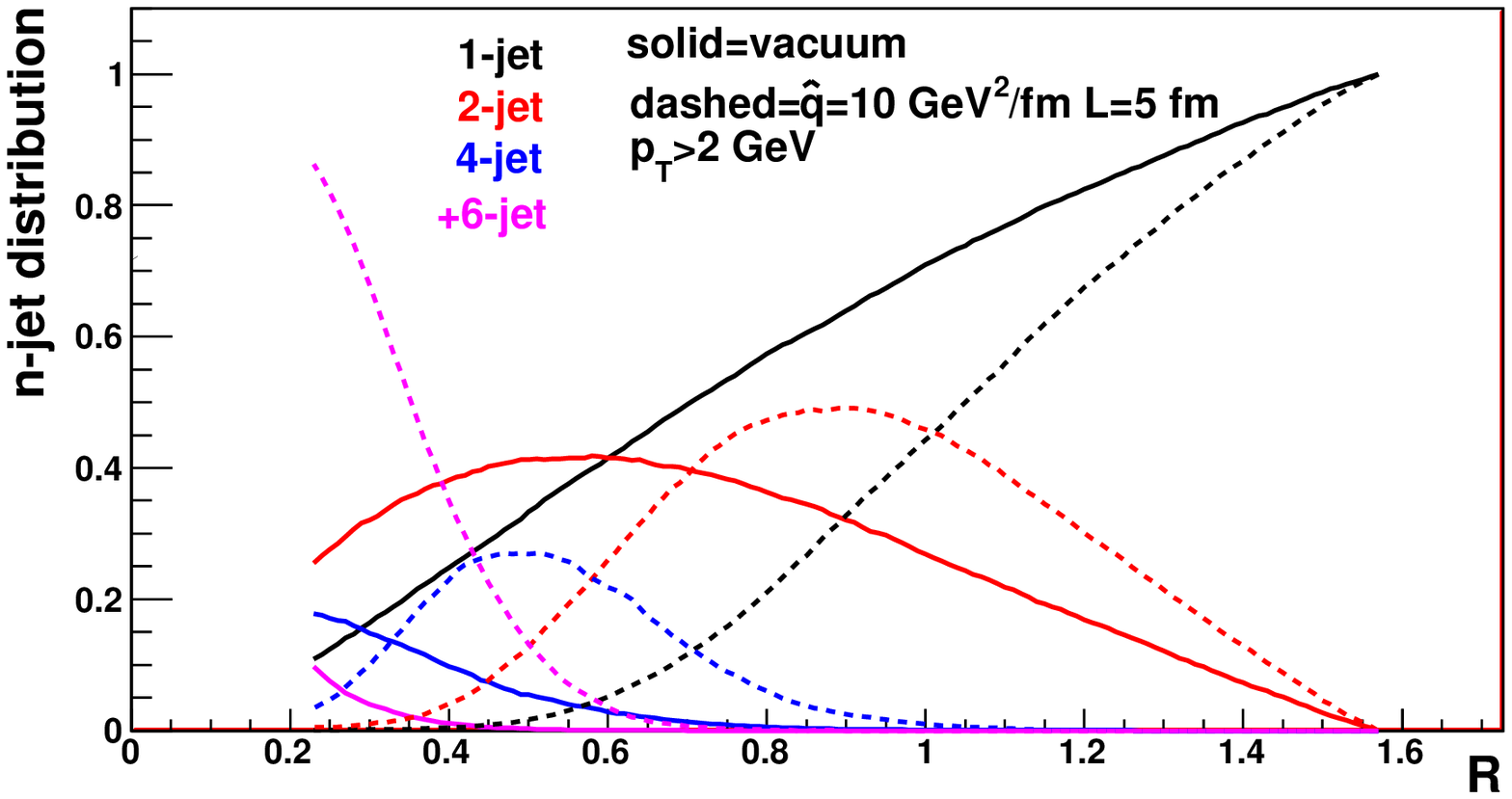}
\caption{Distributions in jet multiplicity for the partonic decay products of a gluon jet of energy 100 GeV, using the anti-$k_T$ recombination 
algorithm, versus the radius of the algorithm. Solid lines refer to results without medium, dashed lines for a medium with $L=5$ fm and $\hat q=10$ GeV$^2$/fm. Black/red/blue/magenta lines refer to 1-,2-,4- and $n\ge 6$-jet distributions respectively. The upper plot contains all reconstructed jets without any cut on their $p_T$, while the lower plot shows the results for jets with $p_{T}^{j}>2$ GeV. See the text for explanations.}
\label{fig:8}      
}

Note that all the extracted conclusions about jets can only be taken as illustrative of the effects of quenching in our model. Further ongoing work on jet reconstruction within a heavy ion environment using different jet-definition algorithms and state-of-the-art background subtraction techniques \cite{bgst} is required to further substantiate these conclusions.

\section{Conclusions and outlook}
\label{conclu}

We present a medium modification of the final-state branching process implemented through an additive term in the splitting functions following the ideas in Ref. \cite{msf}. In order to develop a tool for public use we have modified the parton shower routines within PYTHIA \cite{Sjostrand:2006za} in two ways: first we have included the exact leading order vacuum splitting functions through numerical integrations --- this modification was checked to agree with normal PYTHIA; second we added a term to these splitting functions given by the medium-induced gluon radiation spectra within the BDMPS approximation. This implementation, which is not an official PYTHIA release, is nicknamed Q-PYTHIA and is publicly available at the site \cite{pweb}.

At large jet energies, radiative processes are expected to dominate the modification of the parton shower evolution in a medium. This leading contribution is the one included here. Other contributions which could reveal  to be relevant in different regions of phase space and are not yet taken into account in our formalism, will be considered in the future. These include: i) non-eikonal corrections to the splitting functions; ii) medium-modified color reconnections; iii) energy flow from and to the medium; iv) different hadronization scenarios; v) role of the ordering variable, etc.  

The main aim of the present work is to provide the tools for the upcoming data on reconstructed jets in heavy-ion collisions. These data are starting to appear at RHIC energies and they are expected to be extremely relevant at LHC energies. For this reason we do not seek here any comparison with present experimental data. We have presented, instead, a set of observables to check the feasibility of our approach, in particular to reproduce qualitative or semi-quantitative features expected for in-medium jets. We have shown the suppression of leading particles in fragmentation functions as well as the corresponding enhancement of the hump-backed plateau; the broadening of the angular distributions; and the enhancement of intra-jet multiplicities which corresponds, as expected, to a larger number of splittings. Hadronization tends to diminish the medium effects on soft particles but the expected features remain. Energy-momentum conservation is seen to play a relevant role with respect to naive expectations for some of these quantities. This fact provides further support to the use of Monte Carlo techniques for studying the present problem. 

Additionally, we have presented some examples to illustrate the ability of our formalism in simulating experimentally accessible quantities. In particular, we have presented the broadening of the jet energy and the enhancement of the $n$-jet distributions. Both of them are computed in a very idealized setup, in particular without any background, and should not be considered, in any respect, predictions to be compared with experimental data before a rigorous feasibility analysis is performed. In order to check the degree of convergence of our approach to previous ones, we have also computed the $R_{AA}$ using the same geometry and parameters as in Ref. \cite{Dainese:2004te} obtaining a reasonable agreement with the findings at RHIC.

\section*{Acknowledgments}

We thank N. Borghini, M. Cacciari, G. Corcella, D. d'Enterr\'{\i}a, F. Krauss, J. Rojo-Chac\'on, G. Salam, T. Sj\"ostrand, K. Tywoniuk and U. A. Wiedemann for useful discussions. 
This work has
been supported by Ministerio de Ciencia e Innovaci\'on of Spain under
projects FPA2005-01963, FPA2008-01177 and contracts Ram\'on y Cajal
(NA and CAS); by Xunta de Galicia (Conseller\'{\i}a de Educaci\'on and
Conseller\'\i a de Innovaci\'on e Industria -- Programa Incite) (NA
and CAS); by the Spanish Consolider-Ingenio 2010 Programme CPAN
(CSD2007-00042) (NA and CAS); and by the European Commission grant
PERG02-GA-2007-224770 (CAS).

\appendix
\section{A brief user's guide}
\label{manual}

The program Q-PYTHIA(v1.0), publicly available in \cite{pweb}, is standard PYTHIA(v6.4.18) in all respects (including all defaults), except for a modified version of routine {\tt PYSHOW}  for virtuality-ordered final state radiation. (Note that more recent fortran versions of PYTHIA \cite{Sjostrand:2006za} 6.4.19 or 6.4.20 do not modify {\tt PYSHOW} but for the possibility of forbidding final state radiation of specific partons.) The files to be found in \cite{pweb} are the following:

\begin{enumerate}

\item {\bf q-pyshow.1.0.f}: it contains modified {\tt PYSHOW} plus several additional routines for computing the medium-modified splitting functions, including two routines {\tt QPYGEO} and {\tt QPYGIN} describing the geometry and to be modified by the used --- see below. It is to be linked with a main file, a version of PYTHIA without its standard {\tt PYSHOW}, and the CERNLIB libraries \cite{cernlib}.

\item {\bf q-pythia.1.0.f}: it contains all PYTHIAv6.4.18 with standard {\tt PYSHOW} replaced by modified {\tt PYSHOW}, plus several additional routines for computing the medium-modified splitting functions, including the geometry routines mentioned above. It is to be linked with a main file and the CERNLIB libraries \cite{cernlib}. This file contains the whole set of routines needed to be linked to a main program. By setting the medium parameters to zero this code reproduces standard PYTHIA --- see Section \ref{montecarlo}.

\item {\bf main-q-pythia.f}: example of main program to make final state radiation on a single gluon. This is the setup considered in most of the plots in the present paper.

\item {\bf qpyrobo.f}: additional routine {\tt QPYROBO} to make Lorentz transformations and boosts from the center of mass system of the hard scattering where PYTHIA performs the final state shower, to the center of mass system of the collision --- see below. It is to be linked with previous routines, if required.

\item {\bf grid-qp.dat}: data file to be located in the same directory.

\end{enumerate}

The Q-PYTHIA release contains some geometry examples included in the routines  {\tt QPYGEO} and {\tt QPYGIN} which can be chosen by the user. At this point, they are taken as indicative and the user is supposed to include his/her own favorite geometry. The only two parameters entering the medium term in the splitting functions are $\omega_c$ and $\hat qL$. For a general medium, these effective parameters are usually computed using the scaling relations \cite{Dainese:2004te,Salgado:2002cd,Renk:2006pk,Renk:2006sx,Bass:2008rv,Armesto:2009zi}
\begin{eqnarray}
\omega_c^{eff}(x_0,y_0,z_0,\tau,\beta_x,\beta_y,\beta_z)=\int d\xi \, \xi \hat{q}(\xi),\nonumber\\ 
\protect[ \hat{q}L ]^{eff}(x_0,y_0,z_0,\tau,\beta_x,\beta_y,\beta_z)=\int d\xi \, \hat{q}(\xi),
\label{eq:scaling}
\end{eqnarray}
where $x_0$, $y_0$, $z_0$ and $\tau$ are the initial position and time to be considered (the initial production point or the point of last splitting) and $\beta_i$, $i=x,y,z$, the components of the corresponding trajectory three-velocity in units of $c$. The linear integrations in Eq. (\ref{eq:scaling}) are to be performed along the trajectory defined by the inputs.

This geometry is implemented through the two routines {\tt QPYGEO} and {\tt QPYGIN} as follows:
\begin{enumerate}

\item In routine {\tt QPYGIN(X0,Y0,Z0,T0)} the user must provide the initial position and time of the parton in the medium (or of the hard scattering). While usually these coordinates are given in the center of mass system of the collisions (as e.g. positions generated by the product of overlap functions as in PQM \cite{Dainese:2004te}), the user must transform them to the center of mass system of the hard scattering --- when applicable ---  using routine {\tt QPYROBO}. As a default, all coordinates and times are set to 0.

\item In routine {\tt QPYGEO(X,Y,Z,T,BX,BY,BZ,QHL,OC)} the user must provide the values of the parameters {\tt QHL}  ($\hat q L$, in GeV$^2$) and {\tt OC} ($\omega_c$, in GeV) for a parton located at {\tt (X,Y,Z,T)} and traveling along the direction defined by the three vector {\tt (BX,BY,BZ)} --- normally by using Eqs. (\ref{eq:scaling}) with a user definition of the local transport coefficient, $\hat q(\xi)$, along a given linear trajectory parametrized by $\xi$. The input values of this routine, {\tt X,Y,Z,T,BX,BY,BZ}, are provided by Q-PYTHIA during the evolution process. I.e. for the first scattering they are taken as the initial ones as defined in  {\tt QPYGIN(X0,Y0,Z0,T0)} and at each step of the evolution, Q-PYTHIA computes the new position after a splitting (through the formation time as explained in Section \ref{montecarlo}).  
Some examples of exact and approximate simple geometries are provided. Note that the position, time and direction of the trajectory are given in the center of mass system of the hard scattering and the medium model is usually given in the center of mass system of the collision. The user may employ routine {\tt QPYROBO} for the transformation.

\end{enumerate}

Further comments on the use of these routines can be found in the corresponding headers. Note that all positions and times are to be given in fm.

Compilation and running have been tested in different platforms and compilers (g77, gfortran) and with different operative systems (MacOS and several versions of linux). Note that in version 1.0.1 linking with CERNLIB is no longer required. The performance is considerably slower than standard PYTHIA, noticeably for large energies of the showering partons and large medium $\hat q L$ and $\omega_c$. This is due in part to the complexity of the generation of virtualities and momentum fractions from the medium-induced splitting functions, but mainly to the larger amount of branchings that occur. Work to improve the performance, and to include more complex examples of medium models, is under progress.

\end{document}